# Ordeal Mechanisms, Information and the Cost-effectiveness of Subsidies: Evidence from Subsidized Eyeglasses in Rural China

Sean Sylvia, Xiaochen Ma, Yaojiang Shi, Scott Rozelle, and C.-Y. Cynthia Lin Lawell[*]

December 30, 2015

**Abstract:** The cost-effectiveness of policies providing subsidized goods is often compromised by limited use of the goods provided. Through a randomized trial, we test two approaches to improve the cost-effectiveness of a program distributing free eyeglasses to myopic children in rural China. Requiring recipients to undergo an ordeal better targeted eyeglasses to those who used them without reducing usage relative to free delivery. An information campaign increased use when eyeglasses were freely delivered but not under an ordeal. Free delivery plus information was determined to be the most socially cost-effective approach and obtained the highest rate of eyeglass use.

JEL *Codes*: C93, D12, H42, D80

[*] Sylvia: Renmin University of China, 59 Zhongguancun Ave., Beijing 100872, China (e-mail: ssylvia@ruc.edu.cn); Ma (Corresponding Author): Peking University, 112 Shu Wahh Building, 38 XueYuan Road, Haidian District, Beijing 100191, China (e-mail: mxc4068@gmail.com); Shi: Shaanxi Normal University, 620 Chang'an Road West, Xi'an 710119, China (e-mail:shiyaojiang7@gmail.com); Rozelle: Stanford University, 616 Serra St., Encina Hall East Wing Rm. 401, Stanford, CA 94305 (e-mail: rozelle@stanford.edu); Lin Lawell: University of California at Davis, One Shields Ave., Davis, CA 95616 (e-mail: cclin@primal.ucdavis.edu). We

are indebted to Steve Boucher, Pascaline Dupas, and Vivian Hoffmann for detailed comments on previous drafts. We are also grateful to helpful comments from seminar participants at NEUDC, PacDev, Stanford University, University of Southern California, Peking University, Shanghai Jiao tong University, Tongji University. We also acknowledge the great effort of 300 enumerators from the Center for Chinese Agricultural Policy, Chinese Academy of Sciences; Renmin University of China; and students at the Center for Experimental Economics in Education (CEEE) at Shaanxi Normal University. The efforts of Xiaopeng Pang, Junxia Zeng, Jingchun Nie, Lei Wang, Matt Boswell and other field research managers made this study possible. Staff from the Zhongshan Ophthalmic Center, Sun Yat-sen University, gave invaluable eye-related guidance. We are also grateful for financial and technical support from OneSight, Luxottica-China, Essilor and CLSA.

# Ordeal Mechanisms, Information and the Cost-effectiveness of Subsidies: Evidence from Subsidized Eyeglasses in Rural China

## I. Introduction

The cost-effectiveness of programs providing goods and services for free or at subsidized prices is often compromised by the fact that many individuals do not use the goods provided. Many goods, such as insecticide-treated mosquito nets (Cohen and Dupas, 2010), water purification technologies (Ashraf et al., 2010; Dupas, 2014), improved cookstoves (Miller and Mobarak, 2013) and eyeglasses (Glewwe et al., 2012), for example, require active use for benefits to be realized. Low utilization rates, even after goods have been acquired, imply that subsidies are often wasted on goods that go unused. Moreover, the cost-effectiveness of subsidization programs can also be undermined if subsidies displace pre-existing demand at market prices.

The cost-effectiveness of subsidization programs can be improved by (a) targeting to individuals most likely to use them and least likely to purchase them at market prices (on the extensive margin); and (b) increasing usage among those acquiring a good (on the intensive margin). Although it has been argued that charging a positive price (or "cost-sharing") can accomplish both tasks, empirical evidence suggests that charging even a modest price for some products can significantly dampen demand and screen out a significant number of individuals who would otherwise use the products had they been provided for free, thus compromising the primary policy goal of increasing usage (Cohen and Dupas, 2010; Ashraf et al., 2010).[1]

In this paper, we explore two non-price policy approaches aimed at improving the cost-effectiveness of programs providing goods for free with the ultimate goal of increasing usage. The first approach, applying a so-called ordeal mechanism to product distribution, aims to improve cost-effectiveness by targeting products to individuals who will use the product (Dupas et al., 2013). The second approach, providing information through an information campaign, attempts to increase usage among recipients of the product by addressing misinformation that contributes to low rates of use. Through a field experiment in which we distributed eyeglasses to myopic children in rural China, we test the impact of these two approaches on uptake, overall use and cost-effectiveness. Because subsidies and information (or social marketing) campaigns are commonly implemented simultaneously (Ashraf et al., 2013), we also test how the effects of an ordeal and information interact.

Although traditionally used to target government transfers, such as welfare and unemployment (Nicols et al., 1971; Nicols and Zeckhauser, 1982; Besley and Coate, 1992; Alatas et al., 2013), ordeal mechanisms have more recently been examined as a way to distribute subsidized goods in developing countries (see Dupas et al., 2013). Ordeal mechanisms attempt to encourage targeted individuals to self-select into programs by requiring that applicants undergo an ordeal, such as a time-consuming application procedure or traveling to redeem a voucher.[2] In the case of subsidized goods, imposing an ordeal for individuals to acquire the product can improve program cost-effectiveness by selecting in individuals who would likely use a good and selecting out individuals unlikely to use a good, as well as those more likely to purchase a good at market prices (Dupas et al., 2013).[3] A potential cost of ordeal mechanisms, however, is that (similar to price instruments) they can screen out individuals who would have otherwise used the product. If the ordeal mechanism reduces access to those that would have otherwise used a product, then it could be counter-productive to the policy goal of increasing overall use. Moreover, because ordeal mechanisms work by imposing a deadweight loss, social cost-effectiveness may not improve if this loss is not compensated by a reduction in non-users acquiring a good.

---

[1] In addition to screening out those unlikely to use a product, charging a price may increase use through sunk-cost effects (the psychological effect whereby individuals are more likely to use a good because they paid for it, independent of the intrinsic welfare the good brings—Thaler, 1980) or by signaling that a product is high quality. There is little evidence, however, that cost-sharing increases utilization (Cohen and Dupas, 2010; Ashraf et al., 2010).

[2] Note that although vouchers are in fact a relatively common way of distributing subsidized goods – such as agricultural input subsidies, for example (Jayne and Rashid, 2013) – we are aware of no studies that examine these policies as a self-selection mechanism.

While an ordeal mechanism can improve cost-effectiveness through targeting, providing information about a good through an information campaign can improve cost-effectiveness by increasing the rate or intensity of usage among recipients of the product. Applied to experience goods, providing information on the true benefits of a good can increase uptake when beneficiaries either have little prior information or have priors that undervalue a good (Kremer and Glennerster, 2011). Information campaigns also can increase uptake by reducing the perceived or real costs of using a good, for example by reducing stigma associated with product use (Brent, 2010). Apart from providing new information about a good, a second way information campaigns may increase uptake is by increasing the *salience* of a good—a potentially important channel if low utilization is driven by limited attention rather than lack of knowledge (Kremer and Glennerster, 2011).

---

[3] While the objective of applying an ordeal mechanism to the distribution of a subsidized product is to target individuals based on perceived benefit, traditional ordeal mechanisms applied to transfer programs target individuals based on cost. Because the utility cost imposed by the ordeal is different for the rich than the poor because poor people may be more liquidity-constrained but less time-constrained than the rich, requiring that applicants undergo such an ordeal can deter the relatively wealthy from applying for benefits thus reserving government resources for the needy. In the case of transfers, ordeal mechanisms have been shown to be both cheaper (eliminating the need for costly or impossible collection of data needed for means testing, for example) and lead to more efficient targeting (Alatas et al., 2013).

Given that the effectiveness of the ordeal mechanism in screening out the "right" individuals depends on perceptions of the benefits and costs of using a product at different points in time, a simultaneous information campaign aimed at changing these perceptions could have important effects on the performance of the ordeal. For instance, an information campaign may induce more individuals to undertake an ordeal to acquire a good; however, cost-effectiveness will be reduced if these individuals do not also subsequently use the good acquired. Moreover, if an information campaign has a greater effect on ex-post usage than on uptake under an ordeal mechanism, an ordeal may screen out individuals who would have used a good if they had the good and were also informed.

Through a large-scale field experiment involving 252 schools in rural China, we study the effects of an ordeal mechanism and information campaign in the context of a program distributing eyeglasses to myopic school children. We first randomly assigned schools to one of three distribution groups: a *Free Delivery* group in which eyeglasses were delivered to schools; an *Ordeal* group in which students and their family members were required to travel to the local county seat to redeem vouchers for otherwise free glasses; and a standard of care control or *Market* group in which students were only given eye examinations and a prescription to obtain eyeglasses in the existing eyeglass market. Half of the schools within each of these distribution groups were then randomly allocated to receive an *Information Campaign*. Through this experimental design, we compare the effects of each type of distribution scheme, and their interaction with information, on eyeglass ownership and use in the short term (one month after initial distribution) and long term (seven months after initial distribution).

A program to distribute subsidized eyeglasses is a useful setting to study the questions that we address for at least five reasons. First, uncorrected refractive errors (such as myopia) represent a substantial burden in developing countries, generally, and in China, in particular (Maul et al., 2000; Murthy et al., 2002; He et al., 2007; Resnikoff et al., 2008; Ma et al., 2014). Second, even though eyeglasses can fully correct vision loss due to myopia, they are utilized at low rates in developing countries—likely due to a lack of information (Li et al., 2010; Yi et al., 2015), misinformation (Congdon et al., 2008; Li et al., 2010), or stigma (Castanon Holguin et al., 2006; Odedra et al., 2008). Third, like many health products, eyeglasses are an experience good whose use is subject to both positive and negative learning. Experiential learning may be particularly important in determining demand in places where the quality of medical advice is low, such as rural China (Fischer et al., 2014; Bai et al., 2014; Sylvia et al., 2015). Fourth, because eyeglasses are generally able to fully correct reduced vision due to myopia, an individual's degree of myopia prior to the program provides an indicator for each individual's potential welfare improvement from the use of eyeglasses, a convenient feature for analyzing cost-effectiveness. Finally, from a policy perspective, eyeglasses are costly to provide at scale, warranting the development of strategies to improve the cost-effectiveness of their provision.

We emphasize two main findings. First, without the information campaign, requiring an ordeal modestly improved targeting efficiency. The ordeal selected out approximately 12% of individuals that would not have used glasses if freely distributed to students at schools. The use of eyeglasses (as measured though unannounced visits to schools) was as high in ordeal schools as in the free delivery schools suggesting that no individuals were selected out by the ordeal who would have used eyeglasses if freely delivered. Seven months after the initial distribution, "effective coverage" (defined as usage unconditional on ownership (Cohen and Dupas, 2010)) was approximately 12 percentage points higher in the Ordeal and Free Delivery groups relative to the Market group (an increase of nearly 30 percent).

Second, the information campaign was effective in increasing usage when glasses were freely delivered, yet had no effect on usage when glasses were distributed through an ordeal mechanism or when individuals needed to purchase eyeglasses in the market. Providing information increased usage in the free distribution group by approximately 19 percentage points in the short-term and 8 percentage points in the long-term follow-up. We find evidence that the information campaign was most effective in combination with free delivery due (at least in part) to two related effects: (a) the ordeal mechanism selected out of eyeglass ownership those who would be most responsive to information if eyeglasses had been freely delivered; and (b) undergoing an ordeal mechanism had a salience effect on ex-post use which was a substitute for the effect of the information campaign.

Taking these effects together, our results at seven months after the initial distribution show that free delivery with the information campaign achieved the highest overall rate of effective coverage out of the policy options examined. Moreover, this combination of interventions was the most cost-effective, both in terms of "programmatic" costs (implementation costs) and social costs that account for deadweight loss due to the ordeal. Given the patterns of use that we find, the ordeal would have had to screen out approximately 48 percent more individuals without reducing the overall rate of use in order to be as cost-effective as free delivery with information. Requiring an ordeal costly enough achieve this level of screening, though, would likely impose a deadweight loss large enough to fully offset gains from improved targeting. We stress, however, that this specific pattern of results is likely to be context dependent and hinge in particular on the elasticity of voucher redemption with respect to time costs, the initial knowledge of the good in the target population, as well as the learning attributes of the good being subsidized (the potential for positive and negative learning through product use (see e.g., Fischer et al., 2014)).

Our findings contribute to the literature on the subsidized distribution of goods in developing countries. The majority of this literature has focused on how pricing affects demand and subsequent use and uptake (Cohen and Dupas, 2010; Ashraf et al., 2010; Dupas, 2014; Tarozzi et al., 2014; Fischer et al., 2014). By examining how distribution through an ordeal mechanism affects these outcomes, our findings build most directly on the findings of Dupas et al. (2013) who examine the effects of a "micro-ordeal" in the distribution of chlorine solution for water treatment in Kenya. Dupas et al. (2013) find that requiring households to redeem monthly coupons for free chlorine effectively targets chlorine to those who will use it. They use these results to illustrate that having the policy option of a micro-ordeal dramatically expands the set of parameter values for which a social planner would prefer full subsidization over partial subsidization.

We build on Dupas et al. (2013) in three main ways. First, we corroborate their initial findings on the screening effects of an ordeal in the context of a very different type of experience good. This is important because the effects of distribution and pricing policies could depend a great deal on product characteristics. Second, we find evidence that ordeal mechanisms can increase use through a salience effect. Our third contribution is our finding that the performance the ordeal mechanism relative to free delivery depends on the information provided about the good and that free delivery dominates an ordeal mechanism in terms of effective coverage and cost-effectiveness when combined with information. This third contribution is also related to Ashraf et al. (2013), who find that providing information significantly increases the impact of price subsidies on purchases of water treatments.

The remainder of the paper proceeds as follows. Section 2 provides background on the research context regarding uncorrected refractive errors among school-aged children in rural China. Section 3 describes the experiment and data collection. Section 4 discusses the main results for eyeglass acquisition and usage. Section 5 presents cost-effectiveness calculations. The final section concludes.

## II. Uncorrected Refractive Errors: Prevalence, Consequences and Policy Approaches

The World Health Organization (WHO) estimates that 285 million people are visually impaired, 90 percent of whom live in developing countries (Pascolini and Mariotti, 2012). The main cause of visual impairment is uncorrected refractive errors which accounts for 43 percent of the total burden. Approximately 10-15 percent of school-aged children in developing countries have vision problems due primarily to myopia (Maul et al., 2000; Murthy et al., 2002; He et al., 2007; Resnikoff et al., 2008). With minor exceptions, visual impairment in children can be fully corrected by the timely and proper fitting eyeglasses (Esteso et al., 2007; Ma et al., 2015); however, numerous studies in developing countries document exceedingly low rates of eyeglass use (Bourne et al., 2004; Ramke et al., 2007).

The prevalence of uncorrected refractive errors among children in rural China is among the highest in the world (He et al., 2004; He et al., 2007). Recent studies show that about 25% of primary school students have myopia (Yi et al., 2015) but less than one fifth of myopic children own eyeglasses (Ma et al., 2014; Zhou et al., 2014; Wang et al., 2015).[4] Only half of the children owning glasses regularly wear them (Yi et al., 2015; Wang et al., 2015).

The costs of uncorrected refractive errors are high, particularly among school-aged children. Beyond generally impairing quality of life, uncorrected refractive errors can reduce productivity and can hamper performance at school (Resnikoff et al., 2008). Providing children with eyeglasses through school-based programs has been shown to significantly improve their academic performance, having effects on par with education interventions deemed highly successful (Glewwe et al., 2012; Ma et al., 2014). In a context such as China—where academic performance in early years is especially critical to gaining access to subsequent higher quality educational opportunities—there are potentially substantial long-term benefits of the early correction of myopia.

---

[4] Ma et al (2014) and Zhou et al (2014) find that only 16 percent of students in rural public schools in Shaanxi and Gansu who are myopic own glasses; Wang et al. (2015) find a similar rate (18 percent) in Shanghai migrant schools.

Previous research has highlighted several factors contributing to the low utilization of eyeglasses among school children despite these large potential benefits. Lack of vision screening and the affordability of eyeglasses given liquidity constraints are likely significant barriers (Resnikoff et al., 2008). In China, many children are unaware they are visually impaired (Li et al., 2010; Yi et al., 2015) and eyeglasses are a large expense in poor rural areas.[5] But even in countries where there is routine screening and eyeglasses are freely provided, compliance can remain low; suggesting that other factors—such as misinformation and stigma—may also play a role (Resnikoff et al., 2008).

Misinformation has been highlighted as a major reason for low utilization among children in rural China (Li et al., 2010; Yi et al., 2014). There is a common misconception, for example, that wearing eyeglasses further deteriorates vision, likely a result of the correlation between eyeglass use and deteriorating eyesight. It is also commonly believed that eye exercises can prevent and treat myopia, despite any opthamological basis or evidence on their efficacy (Li et al., 2015).[6] Interestingly, stigma appears to be a relatively minor issue in our sample based on baseline student responses to questions about teasing. Only 15 percent of students that wore glasses regularly reported that they had been teased.

---

[5] Data from our baseline survey showed that among children who owned glasses, the median price paid was approximately $55. This represents a large portion of the average (as of 2011) $130 monthly income for rural families in China (China National Statistics Bureau, 2012).

[6] In our baseline survey, around 75% of parents report that they believe wearing eyeglasses will deteriorate one's vision. Around 55% believe that eye exercises can effectively treat myopia.

If lack of awareness and liquidity constraints are, in fact, significant barriers, providing eye screenings and subsidized eyeglasses may be effective policy approaches to address low utilization. If eyeglasses are subject to positive learning once acquired, even a one-off subsidy may be sufficient to significantly increase long-run use (Dupas, 2014). However, eye screening and subsidies alone may be ineffective if there are significant barriers to uptake and use in the form of misinformation, stigma, or limited attention to eye health. A common approach to addressing these types additional barriers is through information (or social marketing) campaigns.

## III. Experimental Design and Data Collection

### A. Sampling

Our experiment took place in two adjoining provinces of western China: Shaanxi and Gansu.[7] In each of the provinces, one prefecture (each containing a group of seven to ten counties) was included in the study. In each prefecture, we obtained a list of all rural primary schools. To minimize the possibility of inter-school contamination, we first randomly selected 252 townships and then randomly selected one school per township for inclusion in the experiment. Within schools, our data collection efforts (discussed below) focused on 4th and 5th grade students. From each grade, one class was randomly selected and surveys and visual acuity examinations were given to all students in these classes.

---

B. Experimental Design

Following the baseline survey and vision tests, schools were randomly assigned to one of the six cells in the 3 by 2 experimental design shown in Table 1.[8] Schools were first randomized into one of three distribution groups: *Free Delivery*, *Ordeal* or *Market* (or control). Half of the schools assigned to each distribution group were then assigned to receive an additional *Information Campaign*. To improve power, we stratified the randomization by county and by the number of children in the school found to need eyeglasses. In total, this yielded a total of 45 strata. Our analysis takes this randomization procedure into account (Bruhn and McKenzie, 2009).

The three distribution groups were as follows:

*Free Delivery:* In the Free Delivery group each student diagnosed with myopia was given a free pair of eyeglasses as well as a letter to their parents informing them of their child's prescription.[9] The child was permitted to select a pair of frames, which were then fit to the proper prescription and delivered to students at schools by a team of one optometrist and two enumerators.

*Ordeal:* In the Ordeal group, each student diagnosed with myopia was given a voucher as well as a letter to their parents informing them of their child's prescription. Their prescription was also printed in the voucher. This voucher was redeemable for one pair of free glasses at an optical store that was in the county seat.[10] To a large extent, the ordeal is simply the cost (in transportation fare and time) associated with travel to this optical shop. The distance from each student's home village and the county seat varied a great deal within our sample, ranging from 1 kilometer to 105 kilometers with a mean distance of 33 kilometers. The vouchers were non-transferable. Information identifying the student, including his/her name, school and county, was printed on each voucher and students were required to present their identification in person to redeem the voucher.

---

[8] After the baseline survey, one of the 252 schools was found to have no children requiring glasses and was excluded from the randomization. The final sample therefore includes 251 schools.

[9] More than 95% of poor vision is due to myopia. The rest is due to hyperopia and astigmatism. For simplicity, we will use myopia to refer to vision problems more generally.

[10] Program eyeglasses were pre-stocked with one chosen optometrist per county.

*Market:* Myopic students in the Market group were given a letter to their parents informing them of their child's myopia status and prescription.

In each of these three provision groups, half of the schools were randomly assigned to receive an information campaign:

*Information Campaign:* The Information Campaign included three components. First, a short documentary-type film was shown to students in classes. Second, students were given a set of cartoon-based pamphlets in classes. Finally, parents and teachers attended a meeting at the school in which they were shown the film and additional handouts were distributed. Each component of the Information Campaign addressed the benefits of wearing glasses and provided information meant to address common misconceptions that lead to inflated perceptions of use costs and contribute to low adoption rates. The Information Campaign specifically addressed the misconceptions that wearing glasses deteriorates vision and that eye exercises can cure myopia.

C. Data Collection

*Baseline Survey and Eye Examination*

A baseline survey was conducted in September 2012. The baseline survey collected detailed information on schools, students and households. The school survey collected information on school infrastructure and characteristics (including distance to county seat). A student survey was given to all students in selected 4th and 5th grade classes. This student survey collected information on basic background characteristics of students and their knowledge of vision health. Students were also asked about their pre-program experience with eyeglasses, including whether they owned eyeglasses, if they regularly wore their eyeglasses, their experience in wearing eyeglasses if they wore them (were they teased) and what their reasons were for not wearing eyeglasses if they did not. Household surveys were also given to these same students, which they took home and filled out with their parents. The household survey collected information on households (e.g., parents' education levels; the value of family assets) that children would likely have difficulty answering, as well as information on parents' knowledge and perceptions of eye problems and the use of eyeglasses.

At the same time as the school survey, a two-step eye examination was administered to all students in the randomly selected classes in all sample schools. First, a team of two trained staff members administered visual acuity screenings using Early Treatment Diabetic Retinopathy Study (ETDRS) eye charts.[11] Students who failed the visual acuity screening test (cutoff is defined by VA of either eye less than or equal to 6/12, or 20/40) were enrolled in a second vision test that was carried out at each school 1-2 days after the first test. This second vision test was conducted by a team of one optometrist, one nurse and one staff assistant and involved cycloplegic automated refraction with subjective refinement to determine prescriptions for children needing glasses.[12] After being fitted with eyeglasses, corrected visual acuity (with eyeglasses) was tested again to make sure both eyes could be improved to 6/12.

*Acquisition and Use of Eyeglasses*

Our analysis focuses on two key variables: eyeglass ownership and use. Acquisition is defined by ownership, i.e., a binary variable taking value of one if a student owns a pair of eyeglasses at baseline or acquired one during the program (regardless of the source). Acquisition includes both self-purchase and voucher redemption (in the case of the Ordeal group). Ownership is one for all students in the Free Delivery group.[13]

---

[11] ETDRS charts are accepted as the worldwide standard for accurate visual acuity measurement (Camparini et al., 2001).

[12] A cycloplegic refraction is a procedure used to determine a person's degree of myopia (refractive error in a more strict terminology) by temporarily paralyzing the muscles that aid in focusing the eye. It is often used for testing the vision of children who sometimes subconsciously accommodate their eyes during the eye examination, making the results invalid.

Use is defined by whether a student wears his or her glasses. Use was measured through unannounced checks to observe whether students were wearing glasses in class. However, because observation of whether glasses were worn at a particular point in time may over- or underestimate the level of regular use, we supplement this measure with self-reported survey responses of whether children wore eyeglasses and whether they wear them regularly outside of class.

Two rounds of unannounced checks were conducted to collect information on the ownership and use of eyeglasses. A short-term check was done in early November 2012 (approximately one month after glasses and vouchers were distributed). In this round, a team of two enumerators made unannounced visits to each of the 251 schools and recorded the number of students in sample classes that were wearing their glasses. To be counted as wearing glasses, the student had to have been observed with their glasses on. After this count was finished, students that had been diagnosed with myopia in baseline eye examination were given a short survey that included questions about whether they owned eyeglasses, how they acquired them, and how often they wore them. In order to check with students that said that they owned glasses actually did, we asked to see the glasses (if they were with the student at school—e.g., if they were in his/her desk or backpack). If the glasses were at home, we followed up with phone calls to the parents.

A second, long-term, check was conducted alongside an endline survey in May 2013 (seven months after eyeglasses and vouchers were distributed and training was conducted). In line with the first round of unannounced checks, a team of two enumerators was sent into schools in advance of the rest of the survey team to conduct classroom checks. In this second round of checks, enumerators were given a list of the children diagnosed with myopia to record individual-level information on their usage of glasses.

---

[13]  We do not adjust ownership for lost or broken eyeglasses during the program because our focus is on cost-effectiveness and costs for these eyeglasses had already been incurred. By the long-term follow-up, the reported rate of lost and broken eyeglasses was also small (5.6% in the Free Delivery group and 4.6% in the Ordeal group; the difference is not significant at the 10% level).

D. Descriptive Statistics and Balance

Of the 19,934 students given eye examinations at baseline, 3,177 (16%) were found to require eyeglasses. Only these students are included in the analysis sample.

The baseline characteristics of the students in our sample are shown in Table 2. The first two columns show the mean and standard deviation in the Free Delivery, No Information Campaign group. Columns 3-7 show coefficients estimated by regressing each of the baseline characteristics on a vector of indicators for the other treatment arms and indicators for randomization strata. Column 8 reports the p-value from test of a null hypothesis that these coefficients are jointly zero. Only three of the 55 coefficients tested are significantly different from zero and none of the joint tests are rejected at conventional levels suggesting good overall balance across experimental groups.

On average, only 18.9% (601 out of 3177) of students who needed glasses had them at the time of the baseline (Row 8). In line with findings in the literature, lack of awareness and misinformation were prevalent and may play a role in explaining this low rate. For example, less than half of myopic children knew they were myopic (Row 9). Nearly 75% of parents believed that wearing eyeglasses could deteriorate one's vision (Row 10). Generally, the education level of parents was low: only 22.6% of children had at least one parent who had attended high school (Row 3). Nine percent of students in the sample had parents who had both migrated elsewhere for work (Row 4).

Attrition at the short and long term visits was limited (Online Appendix Table 1). Only 19 out of 3,177 students could not be followed in the short term and 127 (4%) could not be followed in the long term. There is slightly higher short-term attrition in the Market, No Information Campaign group; however the magnitude is small (1.7 percentage points).

## IV. Results: Acquisition, Ownership and Usage

### A. Acquisition and Ownership

We begin our analysis by comparing ownership in the short and long term across the experimental groups. To do so, we use OLS to estimate the regression

$$y_{ijt} = \alpha + \beta_I Info_j + \beta_O Ordeal_j + \beta_{OI} Ordeal_j \times Info_j + \beta_M Market_j + \beta_{MI} Market_j \times Info_j + X_j'\gamma + \varepsilon_{ijt}, \quad (1)$$

where $y_{ij}$ is a binary indicator for the eyeglass ownership of child $i$ in school $j$ in wave $t$ (short-term or long-term). $Info_j$ is a dummy variable indicating schools receiving the Information Campaign; $Ordeal_j$ is a dummy variable indicating schools in the Ordeal group; and $Market_j$ indicates schools in the Market group. The coefficient $\beta_I$ captures the effect of the information campaign in the Free Delivery group; $\beta_O$ compares ownership in the Ordeal group to that in the Free Delivery group; and $\beta_M$ compares ownership in the Market group to that in the Free Delivery group. The coefficients on the interaction terms, $\beta_{OI}$ and $\beta_{MI}$, give the additional effect of the Information Campaign in the Ordeal and Market groups relative to the effect of the Information Campaign in the Free Delivery group. $X_j$ is a vector of dummy variables for randomization strata. We adjust standard errors for clustering at the school level using the cluster-corrected Huber-White estimator.

The results for ownership are shown in Table 3. The first two columns show estimates from Equation (1) but collapsing data to the school level. The next two columns show the individual level results. Odd-numbered columns show results for the short-term (approximately one month after initial distribution) and even-numbered columns show long-term (seven month) results. Note that all of the children in the Free Delivery group have a value of 1 for ownership in both the short and long term so coefficients show estimated percentage point deviations from full ownership.

The second row in the table shows that, in the short term, the ordeal screened out approximately 15 percent of individuals who would have been given eyeglasses under free delivery. By the long-term wave, ownership increased only by 5 percent in the Ordeal group, suggesting that if individuals were going to redeem their voucher, they did so shortly after receiving one. Seven months after the distribution of vouchers, approximately 88 percent of individuals had redeemed them for free eyeglasses. This high redemption rate is somewhat remarkable given the cost of the ordeal (discussed in more detail below in Section V). Ownership in the Ordeal group was 50 percent higher than in the Market group, which was approximately 40 percent at seven months (row 4). The Information Campaign had no distinguishable effects on ownership in the Ordeal or Market groups (rows 3 and 5).

B. Correlates of Eyeglass Acquisition

In Table 4, we look at correlates of eyeglass ownership in the long term in the Ordeal and Market groups. Although not causal, this analysis provides useful information on (a) how redemption rates are correlated with costs imposed by the ordeal and (b) whether more myopic children and children from poorer households are more likely to redeem their voucher (i.e., to what degree do vouchers target based on need). Additionally, although the Information Campaign had no distinguishable effect on the *levels* of redemption and ownership in the Ordeal and Market groups, it is possible that providing information affected the selection process—i.e. the composition of those selecting into ownership could be different with and without the Information Campaign. To analyze these issues, we estimate linear probability models where the binary dependent variable is equal to one if a student owns eyeglasses in the long-term wave. We estimate correlates separately for each of four experimental groups: Ordeal without the Information Campaign, Ordeal with Information, Market without Information and Market with Information.

Beginning with the model for the Ordeal without Information group (column 1), we find that—conditional on other covariates—the distance to the county seat (a proxy for the cost of the ordeal) is a significant predictor of ownership. Being in the furthest quintile reduces ownership by 21 percentage points compared to the first quintile. We find little association, however, with proxies for degree of need: visual acuity and household asset quintile are largely uncorrelated with ownership (joint p-value = 0.299—though all children in the sample are myopic and in need of eyeglasses). It therefore appears that the cost of going through the ordeal to acquire eyeglasses is the main driver of selection in this group rather than absolute need. We also find a significant relationship with ownership at baseline and whether children think eyeglasses are good looking – possibly indicating students more likely to wear glasses once acquired.

The selection process is significantly different under the Ordeal when combined with the Information Campaign (column 2 - Chow test statistic: 17.22, p-value <0.001). Interestingly, the cost of the ordeal (distance) is, marginally, no longer a significant predictor of ownership (joint p-value = 0.104) and distance is significantly less correlated with ownership than in the model for the Ordeal without Information group.[14] Keeping in mind that these estimates are not causal, providing information appears to damper the effect of the ordeal cost on voucher redemption. This finding contrasts with findings from elsewhere that information is complementary with price subsidies (information increases sensitivity to price (Ashraf et al., 2013)). Households that are classified as informed about eyeglasses at baseline are also (marginally) more likely than uniformed households to redeem their voucher when there is an Information Campaign.[15]

In the Market groups (columns 3 and 4) there is a strong relationship between eyeglass acquisition (purchase) and the degree of myopia (note that higher LogMAR values correspond to worse vision). Household wealth, on the other hand, is still uncorrelated with purchase. Indirectly this suggests that liquidity constraints may not play a significant role in determining eyeglass demand. Rather, the more important role of subsidies in this context may be to facilitate learning through experience (as discussed in Dupas (2014)).

C. Usage

Usage results are presented in Table 5. We estimate two types of effects on usage: usage conditional on eyeglass ownership (shown in panel A) and unconditional usage or "effective coverage" (panel B). For both types of usage rates, we compare across groups using the same specification as with ownership (equation 1), but with measures of eyeglass usage as the dependent variable. We examine effects on three different measures of usage: whether a student was observed wearing eyeglasses during unannounced checks; whether a student self-reported using eyeglasses at all; and whether a student reported that they use their eyeglasses regularly, including outside of class (See Section III.C for details on each). Although the levels of these different measures vary, as can be seen from the results in Table 5, the observed pattern of results is similar. We therefore focus our discussion on the results based on unannounced checks.

---

[14] Running a separate linear probability model (pooling the ordeal without information and ordeal with information groups) interacting distance measures with a dummy variable for the information campaign gives a p-value of 0.003 for a test that coefficients on these interactions are jointly zero. Entering distance linearly yields a similar result.

[15] Information status at baseline was determined by reducing six baseline variables (parent belief that eyeglasses will harm vision, awareness of own myopia status, whether a family member wears eyeglasses, whether eye exercises are believed to treat myopia, and whether it was believed that children were too young for eyeglasses) into an index using the GLS weighting procedure described in Anderson (2008). Those with index scores greater than the median in the full sample are classified as "informed."

For short-term *conditional* usage (Table 5, panel A, column 1), we find the lowest rates in the Free Delivery without Information group (32.9 percent). Adding the Information Campaign to free delivery increases usage substantially (by 18 percentage points, or a 55% increase). Conditional usage is also 14 percentage points higher under the ordeal than under free delivery. The effects of the Information Campaign and of the Ordeal, however, appear to be substitutes as there is essentially no effect of the Information Campaign under the ordeal. Conditional usage under the ordeal with the Information Campaign was approximately 44 percent, or 11.4 percentage points higher than under free delivery. This is significantly less than the 51 percent rate of conditional usage under free delivery with the Information Campaign (p-value: 0.02). The highest rate of conditional usage was in the Market with Information Group at 65.1 percent, though this is not significantly higher than in the Market without Information Group (60.1 percent).

The pattern of long-term results (seven months after initial distribution) is similar to that in the short term, but the gaps in usage between the groups were narrower. By the long-term follow-up, usage increased slightly in the Free Delivery group (from 32.9 percent to 36.8 percent). Long-term usage in the Free Delivery with Information group remained significantly higher (by 7.7 percentage points) than Free Delivery without Information. Usage in the Ordeal group, however, decreased slightly by 3.5 percentage points (from 46.8 percent in the short-term to 43.3 percent in the long-term) and was no longer significantly higher than usage in the Free Delivery group. Usage in the Ordeal group with Information was also not significantly different than the Free Delivery group. Conditional usage in the Market group remained significantly higher than the other groups at 58.5 percent.

Differences across groups in usage unconditional on eyeglass ownership (Panel B) give an indication of the "effective coverage"—the total number of children wearing eyeglasses—obtained under each of the treatments. We find the highest rate of effective coverage with Free Delivery combined with the Information Campaign, both in the short and long term (51.5 percent in the short-term and 44.4 percent in the long term; unconditional usage is the same as conditional usage in the Free Delivery groups since all students own eyeglasses). This is significantly higher than the Ordeal Groups with and without the Information Campaign in the short-term but not in the long-term (effective coverage was 38-40 percent in the Ordeal group with and without the Information Campaign in both the short and long terms). Usage in the Free Delivery group without the Information Campaign is less than unconditional usage in the Ordeal group, but not significantly so. The lowest effective coverage was in the Market groups (25.9 percent without Information in the long term and 28.6 percent with Information in the long term, difference not significant).

Figure 1 plots together the estimates of ownership and unconditional usage (effective coverage) based on unannounced checks for the short term (Figure 1A) and long term (Figure 1B). For each experimental group, the top grey bar is ownership and the lower green bar is unconditional usage. The grey area represents an estimate of waste: This is the proportion of eyeglasses that were distributed under the program but not used (based on observation during the unannounced check).

In both the short and long term there is a similar pattern. Waste is highest in the Free Delivery only group. In this group, around 63 percent of eyeglasses distributed under the program were not being used at the time of long-term unannounced visits. Requiring that recipients undergo an Ordeal reduced waste by 11 percentage points in the long term to 52 percent (approximately the same amount with and without Information). However, this reduction in waste is only slightly larger than the reduction in waste from adding the Information Campaign to Free Delivery, which reduced waste by 7.6 percentage points to 55.6 percent in the long term. Free Delivery combined with the Information Campaign also achieved the highest rate of unconditional usage (effective coverage). We consider these relative reductions further as part of cost-effectiveness calculations below in Section V.

D. Why Does Information Affect Usage Under Free Delivery but not Under an Ordeal?

Above we find that the Information Campaign had a significant impact in the Free Delivery group but not in the Ordeal group. What could have led to this difference in the impact of information on use across the two methods of distribution? In this section we conduct an exploratory analysis to test two potential explanations.

*Hypothesis 1: Ordeal Selection*

The first hypothesis we consider is that those most responsive to the Information Campaign (in terms of increased usage) were screened out by the ordeal mechanism. In other words, those who chose not to redeem their voucher would have been most responsive to information if eyeglasses were delivered to them. Because these individuals were not screened out of eyeglass ownership under free delivery, they were able to increase usage in response to the Information Campaign.

To examine this hypothesis, we estimate the treatment effects of the Information Campaign on eyeglass usage among those in the Free Delivery group who would be *least* likely to redeem their voucher had they been in the ordeal group. To identify this subgroup, we first use the estimated model of the determinants of voucher redemption in the Ordeal group (discussed in section IV.B above and presented in Table 4) to predict a probability of redeeming a voucher for each child in the Free Delivery group. Although the relevant selection model is voucher redemption with the Information Campaign, we also predict probabilities of redemption *without* the Information Campaign for comparison. We then use these predicted probabilities to classify Free Delivery children into low, medium and high probability subgroups. Finally, we estimate the treatment effect of the Information Campaign within each of these subgroups. To compute standard errors, we bootstrap the entire procedure.

The results of this analysis are presented in Table 6. Using the model of voucher redemption with the Information Campaign (panel B), we find that those in the Free Delivery group with a probability of voucher redemption in the lowest tercile were most responsive to the Information Campaign. We estimate that the Information Campaign increased eyeglass use in this subgroup by 17.9 percentage points. Estimated effects for the medium and high probability groups are not different from zero (point estimates are -3.2 and 3.4 percentage points, respectively). The power of this test is low, however, and the effect of information in the lowest tercile is not statistically larger than in the medium (p-value=0.101) or high probability subgroups (p-value=0.285). Nevertheless, the much larger and significant point estimate for the low probability group lends some support to the hypothesis that selection out of eyeglass ownership under the ordeal mechanism played a role in the different effects of information in the Ordeal and Free Delivery groups.

Using the model of voucher redemption *without* the Information Campaign we find the effects of Information on usage are concentrated among those with a medium probability of redemption (panel A). This suggests that it was not the selection in response to the ordeal itself that reduced the effects of information on ex-post use in the Ordeal group, but rather selection in response to the Ordeal and Information Campaign combined.

*Hypothesis 2: Information and Ordeal Salience*

A second hypothesis for why the Information Campaign affected usage in the Free Delivery group but not in the Ordeal group is that undergoing the ordeal, in itself, increased ex-post use through a channel similar to the channel through which information increased use in the Free Delivery group. Therefore, because this mechanism had already been "exploited" in the Ordeal group, providing information had little additional effect (i.e. the two effects were substitutes). In the Free Delivery group, on the other hand, providing information was able to affect use through this mechanism.

One possible such mechanism is an increase in the salience of eyeglasses. Results from previous studies suggest that information campaigns often work, not by providing new information, but by increasing the salience of a good (Kremer and Glennerster, 2011). The act of undergoing an ordeal could plausibly also increase salience. If salience is an important mechanism for increased usage and salience from these two sources are substitutes for one another, information will have little effect when an ordeal is in place (and vice versa).

To examine the potential role of salience, we classify the sample by "knowledge status" (that is, is one informed or uninformed) at baseline and test the effects of the ordeal and information campaign within these subgroups. As above, we classify a household as "informed" at baseline if they have a baseline information index score above the median and "uninformed" if below the median. If effects operate through increased salience, there should be little effect on uninformed households (increasing the salience of a good thought to be of little value should not increase its use). Alternative mechanisms, such as the Information Campaign affecting use by updating beliefs (i.e., participants in the Information Campaign learned something new about the need to wear eyeglasses) or the ordeal affecting use through a sunk cost effect (Thaler, 1980), would be more likely to increase use among initially uninformed households.

In Table 7 we estimate differences in long-term ownership, conditional usage and unconditional usage between the Free Delivery without information group and the Free Delivery with information group, the Ordeal only group, and the Ordeal with information group; separately for households classified as uninformed at baseline (columns 1-3) and those classified as informed at baseline (columns 4-6).[16] The results are striking. As may be expected, the ordeal selected out approximately 6 percent more uninformed households than informed households (columns 1 & 4 – difference significant, p-value= 0.03). More importantly, however, most of the differences in usage between the groups seem attributable to effects among households informed at baseline. Despite high statistical precision, there are no distinguishable differences in conditional or unconditional use between uninformed households in the Free Delivery-only group and the other groups.

Among informed households, on the other hand, the Information Campaign increased long-term usage by 14 percentage points when glasses were freely delivered. Usage in the Ordeal group was also higher than that in the Free Delivery group (row 2, columns 5 and 6). The effects of the Ordeal and

---

[16] Estimations instead using each variable used to construct the baseline information index are shown in Online Appendix Table 2.

Information Campaign on use, however, were strong substitutes among households informed at baseline. Conditional usage in the Ordeal group with the Information Campaign was not statistically different than in the Free Delivery group with the Information Campaign (p-value = 0.164) or than in the Ordeal group without the Information Campaign (p-value = 0.136). Effective coverage in the Ordeal groups with and without the Information Campaign were similar (p-value = 0.235), but coverage in the Ordeal group with Information was 8.6 percent lower than in the Free Delivery group with Information (p-value = 0.071). Although we cannot say with certainty that these effects are driven by salience, they are consistent with the Information Campaign and the Ordeal both addressing limited attention and the combination of the two not increasing salience further.

In sum, we find some support for both mechanisms (selection under the ordeal and the substitution of salience effects of information and the ordeal) playing a role in explaining why there is an effect of the Information Campaign under free delivery but not under the ordeal. Note, however, that with only 10 percent of households selecting out of ownership by the long term in the Ordeal group, the role of selection under the ordeal likely plays less of a role than the substitution of salience effects.

## V. Cost-Effectiveness

In this section, we compare cost-effectiveness under the different combinations of distribution methods and the Information Campaign, largely following the methodologies discussed in McEwan (2012) and Dhaliwal et al. (2013). In doing so, we consider effectiveness both in terms of the number of children wearing eyeglasses and total vision gain, or gains summed over all children found to be wearing eyeglasses.[17] We also present both "programmatic" cost-effectiveness (using the direct monetary program costs to the implementing organization) and social cost-effectiveness calculations. We calculate social costs at the sum of: (a) programmatic costs; (b) the cost of public funds; and (c) costs incurred by households in responding to the policies (including the costs to households in undertaking the ordeal).

In calculating social cost-effectiveness, one issue is whether the cost of eyeglasses should be counted. On one hand, the cost of transfers such as this represent a cost to the implementing organization, but not to society as a whole (except for the cost of public funds), because the cost to the implementer is offset by the benefit to those receiving the transfer (Dhaliwal et al., 2013). On the other hand, beneficiaries may not value eyeglasses at the cost incurred by the implementer (but their average valuation will still not be zero). Indeed, this is a main motivation for subsidizing goods in the first place. As a compromise, we present social cost-effectiveness calculations both including the cost of eyeglasses as well as calculations that exclude the cost of eyeglasses for those found to be wearing eyeglasses (but still including the cost of public funds corresponding to programmatic eyeglasses purchases).

To account for estimation errors, we use Monte Carlo simulations to calculate how often each policy (combination of distribution methods and the Information Campaign) is likely to be the most cost-effective of those examined (an approach proposed by Evans and Popova (2016)). We use this approach to simultaneously take into account errors for three different effect estimates that serve as inputs to the cost-effectiveness analysis: unconditional use (effective coverage), voucher redemption in the Ordeal groups, and self-purchases of eyeglasses. Each simulation draw randomly selects simulated values for each effect from normal distributions with means and standard deviations from the original coefficient estimates. For each type of cost-effectiveness calculation, we report the percentage of 100,000 draws for which each policy is the most cost-effective.

---

[17] In our calculations we use the number of children found to be wearing eyeglasses in the long-term unannounced check.

Table 8 presents the results.[18] This table shows costs and benefits of each experimental group relative to the market group without the information campaign. The key finding is that the combination of Free Delivery with the Information Campaign is the most cost-effective in terms of both programmatic and social costs (and both including and excluding full eyeglass costs). In this group, we calculate the programmatic cost of one additional bespectacled child to be 1,714 yuan (or $272 at an exchange rate of 6.3 RMB/USD—row 12, column 5). The point estimate for the relative social cost is 1,608 yuan ($255) per child including full eyeglass costs and 947 yuan ($150) excluding eyeglass costs for wearers (rows 14 and 16, column 5). For each of these three types of costs, Free Delivery with the Information Campaign was the most cost effective policy in more than 90 percent of simulation draws.

The second most cost-effective policy (based on the simulation) is the Ordeal *without* the Information Campaign. We calculate programmatic costs for this group to be 2,713 yuan ($431) per additional child (row 12, column 2), or 58 percent higher than the Free Delivery with Information group. In terms of relative social costs—which take into account costs incurred by households in undergoing the ordeal—the Ordeal group without Information is even less cost effective (the cost per additional child (row 14, column 2) is 76 percent higher than the Free Delivery group with Information). The results are similar including and excluding full eyeglass costs (row 16, column 2).

---

[18] See table notes for further details about these calculations.

Accounting for the gain in visual acuity of those ultimately wearing eyeglasses (calculated as the difference between corrected visual acuity—vision with eyeglasses—and presenting visual acuity—vision without eyeglasses), the differences between the cost-effectiveness of the Free Delivery with Information policy and that of the Ordeal without Information policy become even larger (rows 18-23). Programmatic costs are 3,126 yuan ($496) per line of visual acuity improvement (measured in LogMAR units) in the Free Delivery with Information group (row 18, column 5) and 67 percent higher in the Ordeal group without Information (5,225 yuan ($829) per line – row 18, column 2). Social costs per line of improvement are more than 86 percent more expensive in the Ordeal group without Information compared to Free Delivery with Information (rows 20 and 22).[19]

A remaining question is whether the ordeal could have been more cost-effective than free delivery with information if it had screened out more non-users. Given the effects that we find, the ordeal would have needed to screen out around 48 percent of all myopic children—without reducing the number or composition of children using glasses—to achieve the same cost-effectiveness as free delivery with information.[20] Our estimates of how redemption falls with distance to the county seat imply that requiring a travel distance of 181 km would be needed to achieve that amount of screening. This would impose a cost on households of around 648 yuan, more than if they traveled to the county seat and purchased their own. Of course this calculation makes several assumptions, but the point is to illustrate that, in our context, the amount of deadweight loss that would need to be imposed to screen out enough individuals is likely so large that the ordeal could never be more (socially) cost-effective than free delivery with information.

[19] One caveat to these calculations is that they are sensitive to assumptions on the programmatic cost of eyeglasses. Because eyeglasses were donated for the purposes of this program, we use the market cost of eyeglasses (350 yuan) collected as part of the baseline survey (following Dhaliwal et al. (2013)). If the program were scaled up, however, eyeglasses could conceivably be purchased more cheaply if bought in bulk. We calculate that if the programmatic cost of eyeglasses could be reduced to 124 yuan per pair (but the cost of eyeglasses purchased in the market remained at 350 yuan), the Free Delivery *without* Information policy would be as cost-effective as Free Delivery *with* Information (but the usage rate would still be lower).

[20] The total cost of eyeglasses in the Ordeal group would need to be reduced to 90,375 yuan, which corresponds to a redemption rate of 52.5 percent given a cost of 350 yuan for one pair of eyeglasses, for social cost-effectiveness in terms of total visual acuity improvement to be the same as the Free Delivery with Information group. Excluding the direct cost of eyeglasses for wearers (counting eyeglasses as a transfer), the Ordeal would need to screen out 33.6%.

## VI. Discussion and Conclusion

This paper provides new evidence to inform the design of policies distributing subsidized goods in low and middle income countries. In the context of a program distributing subsidized eyeglasses to myopic children in rural China, we compare alternative distribution policies in terms of cost-effectiveness and in terms of the overall rate of use achieved.

We find that an ordeal mechanism requiring beneficiaries to travel to their county seat to redeem vouchers for free eyeglasses improved cost-effectiveness relative to a policy of free delivery by screening out individuals who would not have used eyeglasses if freely delivered. The ordeal also did not reduce the overall rate of usage of eyeglasses compared to free delivery. This finding echoes research by Dupas et al. (2013) who find a similar result for a program distributing water treatment in Kenya.

We also find, however that – when combined with an information campaign designed to provide information on the benefits of eyeglasses and address misinformation about the use of eyeglasses – free delivery outperforms the ordeal mechanism (with and without an information campaign) both in terms of cost-effectiveness and the overall rate of use achieved. One reason for this is that the information campaign increased eyeglass use under free delivery but not under the ordeal mechanism. We present evidence that this is due to the ordeal selecting out individuals most responsive to the information campaign and also to salience effects of the ordeal and information campaign being substitutes.

In presenting these results, we stress that, as with related studies, our particular findings are likely specific to context. For example, an ordeal mechanism would likely perform better in contexts where the elasticity of voucher redemption to time costs is higher. The effects of an information campaign—and how they vary under free delivery or an ordeal mechanism—will also likely depend on whether the target population is initially more or less informed about a good. Results will also likely differ for goods with different attributes, in particular goods with different potential for positive and negative learning over time.

Overall, our results show that there is substantial room for the effectiveness and cost-effectiveness of policies distributing subsidized goods to be improved by adjusting additional design features, such as distribution modes and accompanying information or social marketing campaigns, in addition to (or in combination with) adjusting subsidy levels.

**Figures and Tables**

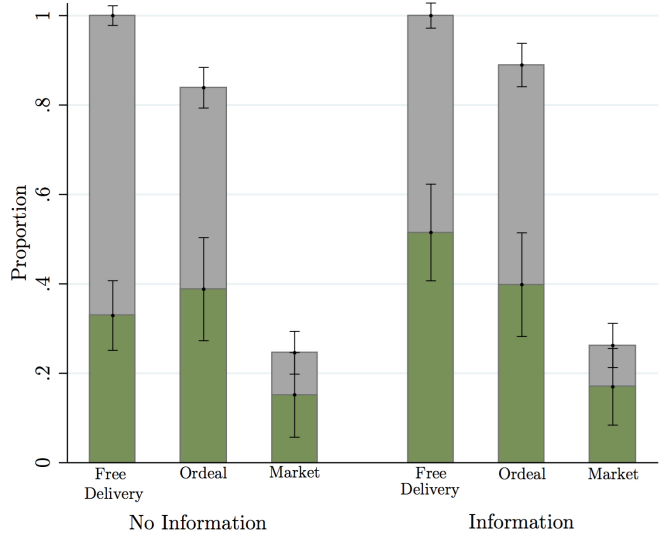

A. Short-term

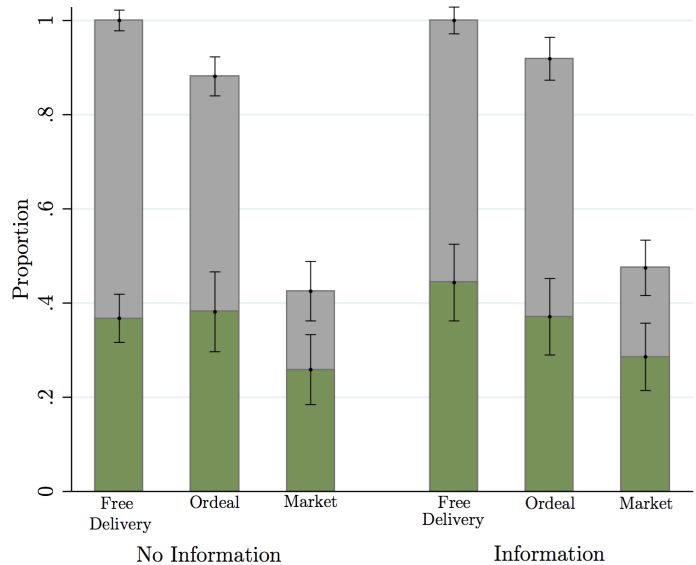

B. Long-term

**Figure 1: Acquisition and Usage by Experimental Group**

*Notes*: Gray bars show rate of eyeglass ownership and green bars show usage rate based on observed use during unannounced visits. Error bars represent 95% confidence intervals. Short-term visits (Figure 1A) were done one month after initial distribution and long-term visits (Figure 1B) were done seven months after initial distribution.

**Table 1: Experimental Design**

| | Distribution Group | | | |
| | Free Delivery | Ordeal | Market | Total |
|---|---|---|---|---|
| No Information Campaign | 42 (527) | 41 (492) | 42 (510) | 125 (1,529) |
| Information Campaign | 42 (626) | 42 (496) | 42 (526) | 126 (1,648) |
| Total | 84 (1,153) | 83 (988) | 84 (1,036) | 251 (3,177) |

*Notes:* Table shows the distribution of schools (myopic students) across experimental groups. In the Free Delivery group, eyeglasses were delivered to schools; in the Ordeal group, vouchers for free eyeglasses were redeemable in the county seat; the Market group only received eyeglass prescriptions. The Information Campaign focused on conveying the benefits of eyeglasses and addressing common misconceptions concerning the use of eyeglasses. One school in the Ordeal, No Information cell was found during the baseline survey to not have any myopic students and was therefore excluded from the randomization.

**Table 2: Descriptive Statistics and Balance Check**

| | | Free Delivery, No Information | | Coefficient (standard error) on: | | | | | | |
| | | Mean | SD | Free Delivery, Information | Ordeal, No Information | Ordeal, Information | Market, No Information | Market, Information | Joint Test P-value | Observations |
| | | (1) | (2) | (3) | (4) | (5) | (6) | (7) | (8) | (9) |
|---|---|---|---|---|---|---|---|---|---|---|
| (1) | Male (0/1) | 0.480 | 0.500 | 0.005 (0.029) | 0.010 (0.029) | -0.002 (0.028) | 0.050 (0.031) | 0.001 (0.031) | 0.500 | 3177 |
| (2) | Grade 5 (0/1) | 0.611 | 0.488 | -0.014 (0.031) | -0.002 (0.034) | -0.004 (0.030) | -0.005 (0.032) | -0.036 (0.031) | 0.829 | 3177 |
| (3) | At least one parent has high school education or above (0/1) | 0.226 | 0.419 | -0.005 (0.025) | -0.058* (0.031) | -0.031 (0.028) | -0.017 (0.027) | -0.028 (0.030) | 0.451 | 3163 |
| (4) | Both parents out-migrated for work (0/1) | 0.092 | 0.289 | 0.003 (0.014) | 0.022 (0.016) | -0.003 (0.017) | 0.009 (0.017) | 0.016 (0.017) | 0.653 | 3147 |
| (5) | Household assets (Index) | -0.057 | 1.290 | -0.104 (0.086) | -0.173* (0.088) | -0.118 (0.101) | -0.127 (0.082) | -0.105 (0.089) | 0.448 | 3032 |
| (6) | Distance to county seat (km) | 33.565 | 22.433 | 2.693 (4.109) | 0.065 (4.419) | -1.991 (4.602) | 5.184 (3.558) | -1.697 (4.080) | 0.365 | 3177 |
| (7) | Visual acuity of worse eye (LogMAR) | 0.629 | 0.202 | 0.002 (0.016) | -0.005 (0.021) | -0.009 (0.016) | -0.003 (0.016) | 0.041** (0.019) | 0.172 | 3177 |
| (8) | Owns eyeglasses (0/1) | 0.188 | 0.391 | 0.014 (0.023) | 0.025 (0.023) | -0.021 (0.022) | 0.021 (0.024) | -0.016 (0.019) | 0.169 | 3177 |
| (9) | Aware of own myopia (0/1) | 0.473 | 0.500 | -0.015 (0.033) | 0.014 (0.033) | -0.011 (0.035) | 0.000 (0.034) | -0.025 (0.035) | 0.894 | 3157 |
| (10) | Parents believe wearing eyeglasses will harm vision (0/1) | 0.747 | 0.435 | -0.002 (0.027) | 0.012 (0.029) | 0.015 (0.027) | 0.002 (0.030) | 0.001 (0.031) | 0.972 | 3011 |
| (11) | Believes eye exercises treat myopia (0/1) | 0.545 | 0.498 | -0.015 (0.035) | -0.016 (0.038) | -0.033 (0.036) | 0.017 (0.039) | -0.010 (0.036) | 0.812 | 3177 |

*Notes*: Data source: baseline survey. The first and second columns show the mean and standard deviation of each baseline characteristic for myopic children in the Free Distribution, No Information group. Severity of myopia is measured by the LogMAR of the worse eye. LogMAR takes value from -0.3 (best vision) to 1.6 (worst vision), with an increment of 0.1 corresponding to a one line change on the vision chart; students with normal vision would have value less than or equal to 0.0. The household asset index was calculated using a list of 13 items and weighting by the first principal component. Distance is the distance from the school to the county seat. Columns 3 through 7 show coefficients and standard errors (in parentheses) from a regression of the characteristic on other five treatment dummies, controlling for randomization strata. Standard errors are clustered at the school level. Column 8 shows the p-value from a Wald test that coefficients are jointly zero. All tests account for clustering at the school level.

***Significant at the 1 percent level.
** Significant at the 5 percent level.
*  Significant at the 10 percent level.

**Table 3: Eyeglass Acquisition and Ownership**

|  | Owns eyeglasses: | School Level | | Individual Level | |
|---|---|---|---|---|---|
|  |  | Short-term (1) | Long-term (2) | Short-term (3) | Long-term (4) |
| (1) | Information | -0.005 | 0.002 | -0.004 | 0.001 |
|  |  | (0.018) | (0.018) | (0.014) | (0.014) |
| (2) | Ordeal | -0.149*** | -0.097*** | -0.164*** | -0.119*** |
|  |  | (0.030) | (0.028) | (0.023) | (0.021) |
| (3) | Ordeal×Information | 0.020 | -0.009 | 0.056 | 0.036 |
|  |  | (0.051) | (0.047) | (0.034) | (0.031) |
| (4) | Market | -0.768*** | -0.610*** | -0.755*** | -0.575*** |
|  |  | (0.028) | (0.036) | (0.024) | (0.032) |
| (5) | Market×Information | 0.041 | 0.069 | 0.021 | 0.048 |
|  |  | (0.044) | (0.053) | (0.036) | (0.045) |
| (6) | Observations | 251 | 251 | 3132 | 3054 |
| (7) | Adjusted R-squared | 0.795 | 0.675 | 0.537 | 0.883 |
| (8) | Mean in Free Delivery, No Information Group | 1.00 | | 1.00 | |

*Notes*: Rows (1) to (5) show coefficients on treatment group indicators estimated by OLS using Equation 1. Standard errors clustered at school level are reported in parentheses. Columns (1) and (2) report school-level estimates and columns (3) and (4) report estimates using individual level data. Columns (1) and (3) report estimates for the short-term follow up one month after initial distribution. Columns (2) and (4) report estimates for the long-term follow up seven months after initial distribution. All regressions control for randomization strata indicators.

*** Significant at the 1 percent level.

** Significant at the 5 percent level.

* Significant at the 10 percent level.

**Table 4: Correlates of Eyeglass Ownership at Long-Term Follow-Up**

| | Sample: | Owns eyeglasses at long-term (seven-month) follow up | | | |
| | | Ordeal, No Information (1) | Ordeal, Information (2) | Market, No Information (3) | Market, Information (4) |
|---|---|---|---|---|---|
| (1) | Visual acuity of worse eye (LogMAR) | 0.001 (0.047) | -0.070 (0.044) | 0.237** (0.092) | 0.429*** (0.102) |
| | Distance to County Seat | | | | |
| (2) | Quintile 2 | 0.000 (0.036) | -0.026 (0.043) | -0.051 (0.097) | 0.328*** (0.068) |
| (3) | Quintile 3 | -0.189*** (0.042) | 0.116** (0.054) | 0.035 (0.096) | 0.328*** (0.089) |
| (4) | Quintile 4 | -0.138** (0.052) | -0.088 (0.101) | 0.139 (0.096) | 0.155** (0.059) |
| (5) | Quintile 5 | -0.211*** (0.037) | -0.047 (0.046) | 0.053 (0.114) | -0.048 (0.123) |
| | Household Assets | | | | |
| (6) | Quintile 2 | 0.034 (0.031) | -0.020 (0.026) | -0.075 (0.047) | -0.057 (0.065) |
| (7) | Quintile 3 | 0.041 (0.038) | -0.011 (0.033) | 0.004 (0.061) | 0.040 (0.048) |
| (8) | Quintile 4 | 0.084** (0.040) | 0.018 (0.027) | 0.004 (0.064) | -0.031 (0.056) |
| (9) | Quintile 5 | -0.013 (0.052) | 0.008 (0.033) | -0.013 (0.059) | -0.006 (0.055) |
| (10) | Owns eyeglasses at baseline (0/1) | 0.126*** (0.028) | 0.035 (0.034) | 0.643*** (0.036) | 0.432*** (0.051) |
| (11) | Thinks eyeglasses good looking at baseline (0/1) | 0.082** (0.031) | 0.028 (0.018) | 0.033 (0.099) | 0.128 (0.145) |
| (12) | Informed (0/1) | 0.027 (0.027) | 0.045* (0.025) | -0.005 (0.032) | 0.118*** (0.040) |
| (13) | Male (0/1) | -0.019 (0.037) | -0.026 (0.027) | 0.012 (0.035) | 0.035 (0.050) |
| (14) | Grade 5 (0/1) | 0.031 (0.031) | 0.004 (0.023) | -0.030 (0.037) | 0.037 (0.052) |
| (15) | At least one parent has high school education or above (0/1) | -0.049 (0.041) | -0.021 (0.031) | -0.120*** (0.044) | 0.032 (0.045) |
| (16) | Both parents out-migrated for work (0/1) | 0.055* (0.032) | 0.022 (0.022) | 0.031 (0.064) | -0.031 (0.063) |
| (17) | Constant | 0.873*** (0.051) | 0.997*** (0.086) | 0.263* (0.132) | -0.149 (0.103) |
| (18) | Observations | 434 | 442 | 455 | 475 |
| (19) | Adjusted R-squared | 0.102 | 0.219 | 0.334 | 0.287 |

*Notes*: OLS estimates of a linear probability model with a binary dependent variable that is equal to 1 if the student owned eyeglasses seven months after initial distribution by experimental group. The regressors were measured at baseline. Sample sizes are less than the full sample due to observations missing at least one regressor. Severity of myopia is measured by the LogMAR of the worse eye. LogMAR takes value from -0.3 (best vision) to 1.6 (worst vision), with an increment of 0.1 corresponding to a one line change on the vision chart; students with normal vision would have value less than or equal to 0.0. Distance is measured as the distance from the school to the county seat. The household asset index was calculated using a list of 13 items and weighting using the first principal component. Information status at baseline was determined by reducing six baseline variables (parent belief that eyeglasses will harm vision, awareness of own myopia status, whether a family member wears eyeglasses, whether eye exercises are believed to treat myopia, and whether it was believed that children were too young for eyeglasses) into an index using the GLS weighting procedure described in Anderson (2008). Those with index scores greater than the median in the full sample are classified as informed. Regressions also include county fixed effects (not shown).
*** Significant at the 1 percent level.
** Significant at the 5 percent level.
*　Significant at the 10 percent level.

**Table 5: Eyeglass Usage**

| | | Usage during unannounced check | | Self-reported: Use eyeglasses | | Self-reported: Regularly wears eyeglasses | |
|---|---|---|---|---|---|---|---|
| | | Short-term (1) | Long-term (2) | Short-term (3) | Long-term (4) | Short-term (5) | Long-term (6) |
| **Panel A: Conditional Usage** | | | | | | | |
| (1) | Information | 0.182*** | 0.077* | 0.113*** | 0.079*** | 0.191*** | 0.083*** |
| | | (0.058) | (0.041) | (0.036) | (0.030) | (0.031) | (0.024) |
| (2) | Ordeal | 0.139** | 0.065 | 0.065* | 0.041 | 0.070** | 0.048 |
| | | (0.065) | (0.047) | (0.038) | (0.039) | (0.031) | (0.030) |
| (3) | Ordeal×Information | -0.207** | -0.109 | -0.085* | -0.051 | -0.139*** | -0.050 |
| | | (0.092) | (0.067) | (0.050) | (0.050) | (0.047) | (0.040) |
| (4) | Market | 0.272*** | 0.217*** | 0.142*** | 0.158*** | 0.220*** | 0.150*** |
| | | (0.072) | (0.046) | (0.051) | (0.036) | (0.058) | (0.042) |
| (5) | Market×Information | -0.132 | -0.115 | -0.081 | -0.043 | -0.113 | -0.041 |
| | | (0.101) | (0.074) | (0.063) | (0.048) | (0.077) | (0.060) |
| (6) | Observations | 251 | 2416 | 2253 | 2416 | 2253 | 2416 |
| (7) | Adjusted R-squared | 0.156 | 0.095 | 0.036 | 0.056 | 0.087 | 0.051 |
| (8) | Mean in Free Delivery, No Information Group | 0.329 | 0.368 | 0.712 | 0.642 | 0.139 | 0.198 |
| **Panel B: Unconditional Usage ("Effective Coverage")** | | | | | | | |
| (1) | Information | 0.186*** | 0.076* | 0.106*** | 0.079** | 0.183*** | 0.081*** |
| | | (0.055) | (0.042) | (0.037) | (0.032) | (0.031) | (0.024) |
| (2) | Ordeal | 0.059 | 0.014 | -0.068* | -0.039 | 0.031 | 0.019 |
| | | (0.059) | (0.043) | (0.039) | (0.038) | (0.028) | (0.027) |
| (3) | Ordeal×Information | -0.176** | -0.086 | -0.033 | -0.025 | -0.122*** | -0.043 |
| | | (0.083) | (0.064) | (0.054) | (0.051) | (0.044) | (0.037) |
| (4) | Market | -0.177*** | -0.109*** | -0.499*** | -0.294*** | -0.045* | -0.042 |
| | | (0.048) | (0.038) | (0.039) | (0.037) | (0.027) | (0.027) |
| (5) | Market×Information | -0.168** | -0.049 | -0.075 | -0.020 | -0.146*** | -0.039 |
| | | (0.065) | (0.056) | (0.051) | (0.051) | (0.038) | (0.038) |
| (6) | Observations | 251 | 3054 | 3132 | 3054 | 3132 | 3054 |
| (7) | Adjusted R-squared | 0.295 | 0.089 | 0.263 | 0.121 | 0.085 | 0.045 |
| (8) | Mean in Free Delivery, No Information Group | 0.329 | 0.368 | 0.712 | 0.642 | 0.139 | 0.198 |

*Notes*: Rows (1) to (5) show coefficients on treatment group indicators estimated by OLS using Equation 1. Standard errors clustered at the school level are reported in parentheses. Panel A shows estimated effects on eyeglass use conditional on eyeglass ownership. Panel B shows estimated effects on eyeglass use not conditional on ownership. Columns (1) and (2) report results from "unannounced checks" where enumerators visited schools unannounced and noted which children were wearing eyeglasses in class. The protocol used in the short term unannounced check was for enumerators to only count the number of children wearing eyeglasses. As a result, regressions in column (1) are at the school level. Columns (3) and (4) report results for self-reported use and columns (5) and (6) report results for self-reported regular use of eyeglasses. Odd-numbered columns report estimates for the short-term follow up one month after initial distribution. Even-numbered columns report estimates for the long-term follow up seven months after initial distribution. All regressions control for randomization strata indicators.

*** Significant at the 1 percent level.

** Significant at the 5 percent level.

* Significant at the 10 percent level.

**Table 6: Endogenous Stratification Estimates - Impact of Information on Eyeglass Usage under Free Delivery, by Probability of Undertaking Ordeal**

| | N | β | Bootstrap-SE |
|---|---|---|---|
| Panel A: By probability of redeeming voucher *without* Information Campaign | | | |
| (1) Low | 346 | -0.012 | (0.083) |
| (2) Medium | 331 | 0.200** | (0.085) |
| (3) High | 356 | -0.018 | (0.088) |
| Panel B: By probability of redeeming voucher *with* Information Campaign | | | |
| (1) Low | 316 | 0.179* | (0.097) |
| (2) Medium | 367 | -0.032 | (0.084) |
| (3) High | 350 | 0.034 | (0.095) |

*Notes*: Table reports estimated effects of the information campaign in the free delivery group on the probability a student is wearing eyeglasses at the time of long-term unannounced visits by the estimated probability that a voucher would have been redeemed had the student been in the ordeal group. Panel A reports estimated effects by probability that a voucher would be redeemed without the information campaign and Panel B reports effects by the probability that a voucher would be redeemed with the information campaign. The probability that a voucher would have been redeemed is estimated by modeling voucher redemption in the ordeal group with and without the information campaign and using estimated coefficients to predict probabilities in the free delivery group (voucher redemption model estimates are shown in Columns (1) and (2) of Table 4). Row 1 in each panel shows the effect of the information campaign among students in the free delivery group with an estimated probability of voucher redemption in the lowest tercile of the sample; row 2 shows estimates for the middle tercile; and row 3 shows estimates for the highest tercile. Cluster-adjusted bootstrap standard errors are calculated using 400 iterations of the entire procedure.

*** Significant at the 1 percent level.

** Significant at the 5 percent level.

* Significant at the 10 percent level.

**Table 7: Impacts of Ordeal and Information Campaign Relative to Free Delivery Without Information Campaign, By Information Status at Baseline**

|     |                   | Uninformed at baseline | | | Informed at baseline | | |
|-----|-------------------|-------------|-------------|-------------|-------------|-------------|-------------|
|     |                   | Ownership | Conditional Usage | Unconditional Usage | Ownership | Conditional Usage | Unconditional Usage |
|     |                   | (1) | (2) | (3) | (4) | (5) | (6) |
| (1) | Information       | -0.010 | 0.005 | 0.001 | 0.003 | 0.139*** | 0.141*** |
|     |                   | (0.015) | (0.055) | (0.055) | (0.008) | (0.047) | (0.046) |
| (2) | Ordeal            | -0.139*** | 0.020 | -0.027 | -0.082*** | 0.159*** | 0.120** |
|     |                   | (0.027) | (0.052) | (0.049) | (0.019) | (0.057) | (0.053) |
| (3) | Ordeal×Information | 0.041 | -0.006 | 0.016 | 0.036 | -0.226*** | -0.206*** |
|     |                   | (0.037) | (0.079) | (0.077) | (0.024) | (0.075) | (0.072) |
| (4) | Constant          | 1.004*** | 0.321*** | 0.321*** | 0.998*** | 0.375*** | 0.373*** |
|     |                   | (0.011) | (0.031) | (0.031) | (0.006) | (0.035) | (0.035) |
| (5) | Observations      | 916 | 863 | 916 | 1023 | 994 | 1023 |
| (6) | Adjusted R-squared | 0.115 | 0.074 | 0.080 | 0.058 | 0.095 | 0.094 |

*Notes*: Table shows estimated effects on eyeglass ownership and usage at the long-term follow up relative to the free delivery without information group. Standard errors in parentheses account for clustering at the school level. Conditional usage is estimated using the subsample owning eyeglasses. Columns 1-3 show effects among those being classified as uninformed about eyeglasses at baseline and columns 4-6 show effects for those classified as informed at baseline. Information status at baseline was determined by reducing six baseline variables (parent belief that eyeglasses will harm vision, awareness of own myopia status, whether a family member wears eyeglasses, whether eye exercises are believed to treat myopia, and whether it was believed that children were too young for eyeglasses) into an index using the GLS weighting procedure described in Anderson (2008). Those with index scores greater than the median in the full sample are classified as informed. Estimated effects by each of these variables individually are shown in Appendix Table 2.
*** Significant at the 1 percent level.
** Significant at the 5 percent level.
* Significant at the 10 percent level.

**Table 8: Cost Effectiveness Calculations**

| | | Incremental Amount Relative to Market, No Information Group | | | |
|---|---|---|---|---|---|
| | Market, Information | Ordeal, No Information | Ordeal, Information | Free Delivery, No Information | Free Delivery, Information |
| | (1) | (2) | (3) | (4) | (5) |
| **Panel A. Costs** | | | | | |
| *Programmatic Costs* | | | | | |
| (1) Information Campaign | 14,154 | -- | 14,154 | -- | 14,154 |
| (2) Eyeglasses | -- | 151,536 | 159,191 | 184,450 | 219,100 |
| (3) Eyeglass Distribution | -- | -- | -- | 16,800 | 16,800 |
| *Costs to Households* | | | | | |
| (4) Voucher Redemption | -- | 54,691 | 56,036 | -- | -- |
| (5) Self-Purchase | 6,340 | (78,112) | (87,372) | (70,248) | (65,507) |
| *Cost of Public Funds* | | | | | |
| (6) Cost of Taxation | 2,831 | 30,307 | 34,669 | 40,250 | 50,011 |
| *Total Costs* | | | | | |
| (7) Programmatic | 14,154 | 151,536 | 173,345 | 201,250 | 250,054 |
| (8) Social | 23,325 | 158,422 | 176,678 | 171,252 | 234,558 |
| (9) Social - Excluding Eyeglass Costs for Wearers | 23,325 | 94,708 | 115,918 | 103,006 | 138,154 |
| **Panel B. Benefits** | | | | | |
| (10) Number of Additional Children Wearing Eyeglasses | N.S. | 56 | 52 | 62 | 146 |
| (11) Total Lines of Visual Acuity Improvement (LogMAR) | N.S. | 29 | 29 | 36 | 80 |
| **Panel C. Cost Effectiveness** | | | | | |
| *Number of Children Wearing Eyeglasses (RMB/Child)* | | | | | |
| (12) Programmatic | N.S. | 2,713 | 3,307 | 3,254 | 1,714 |
| (13)     *% Draws Most Cost-Effective* | | 6% | 1% | 0% | 92% |
| (14) Social | N.S. | 2,836 | 3,370 | 2,769 | 1,608 |
| (15)     *% Draws Most Cost-Effective* | | 3% | 0% | 1% | 95% |
| (16) Social - Excluding Eyeglass Costs for Wearers | N.S. | 1,696 | 2,211 | 1,666 | 947 |
| (17)     *% Draws Most Cost-Effective* | | 3% | 0% | 1% | 95% |
| *Lines of Visual Acuity (RMB/LogMAR units)* | | | | | |
| (18) Programmatic | N.S. | 5,225 | 5,977 | 5,590 | 3,126 |
| (19)     *% Draws Most Cost-Effective* | | 3% | 1% | 1% | 95% |
| (20) Social | N.S. | 5,463 | 6,092 | 4,757 | 2,932 |
| (21)     *% Draws Most Cost-Effective* | | 2% | 0% | 2% | 96% |
| (22) Social - Excluding Eyeglass Costs for Wearers | N.S. | 3,266 | 3,997 | 2,861 | 1,727 |
| (23)     *% Draws Most Cost-Effective* | | 2% | 0% | 2% | 96% |

*Notes.* All costs in renminbi (exchange rate as of Sept. 2012 was 6.3 RMB/USD). In the Market, No Informaiton group, the cost of providing eye exams was 130,050 yuan total (3,096 yuan per school) and the total cost to those self-purchasing eyeglasses was 110,823 yuan. The cost of the information campaign was approximately 337 yuan per school excluding fixed costs of producing campaign materials. As eyeglasses were donated for the program, we use the market cost of eyeglasses (following Dhaliwal et al., 2013) as measured in the baseline survey (350 yuan per pair). The cost of delivering eyeglasses was around 400 yuan per school (costs variable with the number of eyeglasses delivered was relatively small). Voucher redemption costs to households include transportation costs to and from county seat (assumed to be 1 yuan per km traveled) and the time costs incurred. Time costs are calculated based on a daily adult wage of 120 yuan for eight hours of work and calculating total time spent as two-way travel time (assuming an average travel speed of 20 km/hour) plus one hour spent recieving eyeglasses. The cost of self-purchase is calculated as the sum of travel costs to the county seat, time costs (calculated as for voucher redemption), and the cost of eyeglasses (also assumed to be 350 yuan). In the absence of good estimates for China, we use a deadweight loss from taxation of 20% of programmatic costs (Auriol and Warlters, 2012). Lines of visual acuity improvement are calculated as the difference between corrected visual acuity (vision wearing correctly fit eyeglasses) and presenting visual acuity (vision wihtout eyeglasses). Effects not significant (N.S.) for the information campaign in the market group. Rows 13, 15, 17, 19, 21 and 23 show the percentage of 100,000 simulation draws for which each policy is the most cost effective. Simulations draw values for the impact of each policy on unconditional use (effective coverage), voucher redemption (ownership in the ordeal groups), and self-purchase of eyeglasses in the market from normal distributions with means and standard deviations of original impact estimates.

**Online Appendix Table 1: Attrition**

| | Missing Short-Term | | Missing Long-Term | |
|---|---|---|---|---|
| | (1) | (2) | (3) | (4) |
| (1) Information | 0.007 | | 0.005 | |
| | (0.006) | | (0.014) | |
| (2) Ordeal | 0.009 | | -0.002 | |
| | (0.008) | | (0.014) | |
| (3) Ordeal×Information | -0.003 | | 0.001 | |
| | (0.011) | | (0.020) | |
| (4) Market | 0.017** | | -0.003 | |
| | (0.007) | | (0.013) | |
| (5) Market×Information | -0.019* | | -0.019 | |
| | (0.010) | | (0.018) | |
| (6) Visual acuity of worse eye (LogMAR) | | 0.014 | | 0.003 |
| | | (0.013) | | (0.020) |
| Distance | | | | |
| (7) Quintile 2 | | 0.002 | | 0.014 |
| | | (0.007) | | (0.013) |
| (8) Quintile 3 | | 0.002 | | 0.007 |
| | | (0.004) | | (0.007) |
| (9) Quintile 4 | | 0.006 | | 0.011 |
| | | (0.008) | | (0.014) |
| (10) Quintile 5 | | 0.002 | | 0.030** |
| | | (0.008) | | (0.014) |
| Household Assets | | | | |
| (11) Quintile 2 | | 0.003 | | 0.006 |
| | | (0.006) | | (0.011) |
| (12) Quintile 3 | | 0.002 | | -0.004 |
| | | (0.007) | | (0.011) |
| (13) Quintile 4 | | 0.003 | | -0.013 |
| | | (0.007) | | (0.011) |
| (14) Quintile 5 | | -0.003 | | -0.010 |
| | | (0.006) | | (0.010) |
| (15) Owns eyeglasses at baseline (0/1) | | -0.003 | | 0.002 |
| | | (0.005) | | (0.011) |
| (16) Aware of own myopia at baseline  (0/1) | | -0.004 | | -0.000 |
| | | (0.005) | | (0.009) |
| (17) Thinks eyeglasses good looking at baseline (0/1) | | 0.005 | | 0.028 |
| | | (0.012) | | (0.027) |
| (18) Parents believe wearing eyeglasses will harm vision at baseline (0/1) | | 0.005 | | 0.005 |
| | | (0.005) | | (0.008) |
| (19) Male (0/1) | | 0.012** | | -0.006 |
| | | (0.005) | | (0.007) |
| (20) Grade 5 (0/1) | | -0.000 | | 0.004 |
| | | (0.004) | | (0.007) |
| (21) At least one parent has high school education or above (0/1) | | 0.006 | | 0.020** |
| | | (0.006) | | (0.009) |
| (22) Both parents out-migrated for work (0/1) | | 0.005 | | 0.014 |
| | | (0.008) | | (0.012) |
| (23) Constant | 0.006* | -0.026* | 0.041*** | -0.000 |
| | (0.003) | (0.015) | (0.010) | (0.020) |
| (24) Observations | 3177 | 2943 | 3177 | 2943 |
| (25) Adjusted R-squared | 0.009 | 0.007 | -0.001 | 0.001 |
| (26) Mean in Free Distribution, No Information Group | 0.006 | | 0.040 | |

*Notes*: Table shows estimated coefficients from a regression of a missing value for eyeglass ownership in the short and long-term waves on treatment indicators and baseline covariates. Regressions in columns 1 and 3 additionally control for randomization strata and columns 2 and 4 control for county fixed effects. Standard errors in parentheses account for clustering within schools. Some observations are not included in regressions reported in columns 2 and 4 due to missing values for a baseline covariate.

*** Significant at the 1 percent level.

** Significant at the 5 percent level.

 *  Significant at the 10 percent level.

**Online Appendix Table 2: Impacts of Ordeal and Information Campaign Relative to Free Distribution Without Information Campaign, By Information Status at Baseline (Index Components)**

| | Ownership (1) | Conditional Usage (2) | Unconditional Usage (3) | Ownership (4) | Conditional Usage (5) | Unconditional Usage (6) | Ownership (7) | Conditional Usage (8) | Unconditional Usage (9) | Ownership (10) | Conditional Usage (11) | Unconditional Usage (12) | Ownership (16) | Conditional Usage (17) | Unconditional Usage (18) |
|---|---|---|---|---|---|---|---|---|---|---|---|---|---|---|---|
| **Panel A: Uninformed at baseline** | | | | | | | | | | | | | | | |
| *Subsample:* | Parents believe wearing eyeglasses harms vision | | | Not aware of own myopia status at baseline | | | No family members wear eyeglasses | | | Believes eye exercises can treat myopia | | | Believes too young for eyeglasses | | |
| (1) Information | -0.005 (0.010) | 0.062 (0.042) | 0.060 (0.042) | -0.006 (0.013) | 0.036 (0.047) | 0.032 (0.047) | 0.001 (0.011) | 0.094*** (0.044) | 0.093** (0.045) | -0.005 (0.011) | 0.097*** (0.045) | 0.096** (0.045) | 0.001 (0.014) | -0.014 (0.058) | -0.013 (0.058) |
| (2) Ordeal | -0.121*** (0.020) | 0.050 (0.046) | 0.004 (0.042) | -0.150*** (0.029) | 0.028 (0.056) | -0.022 (0.052) | -0.099*** (0.020) | 0.078 (0.050) | 0.042 (0.047) | -0.094*** (0.021) | 0.068 (0.050) | 0.031 (0.045) | -0.116*** (0.030) | -0.032 (0.069) | -0.067 (0.065) |
| (3) Ordeal×Information | 0.057** (0.027) | -0.050 (0.068) | -0.022 (0.065) | 0.047 (0.041) | -0.085 (0.083) | -0.059 (0.079) | 0.002 (0.032) | -0.107 (0.072) | -0.098 (0.070) | 0.019 (0.029) | -0.095 (0.069) | -0.081 (0.066) | 0.055 (0.035) | 0.023 (0.092) | 0.043 (0.090) |
| (4) Constant | 1.001*** (0.008) | 0.349*** (0.025) | 0.347*** (0.025) | 1.003*** (0.009) | 0.280*** (0.030) | 0.282*** (0.029) | 0.999*** (0.008) | 0.309*** (0.026) | 0.307*** (0.026) | 1.000*** (0.008) | 0.317*** (0.028) | 0.316*** (0.028) | 0.994*** (0.010) | 0.365*** (0.034) | 0.361*** (0.033) |
| (5) Observations | 1463 | 1397 | 1463 | 1085 | 1022 | 1085 | 1339 | 1276 | 1339 | 1052 | 1009 | 1052 | 636 | 606 | 636 |
| (6) Adjusted R-squared | 0.106 | 0.100 | 0.104 | 0.103 | 0.071 | 0.071 | 0.095 | 0.084 | 0.088 | 0.097 | 0.095 | 0.098 | 0.123 | 0.088 | 0.093 |
| **Panel B: Informed at baseline** | | | | | | | | | | | | | | | |
| *Subsample:* | Parents do not believe wearing eyeglasses harms vision | | | Aware of own myopia status at baseline | | | Family member wears eyeglasses | | | Does not believe eye exercises can treat myopia | | | Does not believe too young for eyeglasses | | |
| (1) Information | 0.004 (0.014) | 0.138** (0.068) | 0.138** (0.069) | 0.004 (0.008) | 0.137*** (0.047) | 0.142*** (0.046) | 0.000 (0.012) | 0.056 (0.047) | 0.058 (0.047) | 0.015 (0.013) | 0.098** (0.047) | 0.103** (0.047) | -0.000 (0.010) | 0.107*** (0.039) | 0.107*** (0.039) |
| (2) Ordeal | -0.059** (0.023) | 0.183** (0.084) | 0.148* (0.083) | -0.062*** (0.014) | 0.135** (0.056) | 0.107** (0.051) | -0.115*** (0.027) | 0.127** (0.056) | 0.064 (0.052) | -0.116*** (0.024) | 0.110* (0.061) | 0.060 (0.057) | -0.099*** (0.020) | 0.125** (0.051) | 0.080* (0.046) |
| (3) Ordeal×Information | -0.038 (0.037) | -0.319*** (0.106) | -0.318*** (0.104) | 0.019 (0.017) | -0.190** (0.073) | -0.180*** (0.068) | 0.081** (0.031) | -0.185** (0.082) | -0.129* (0.077) | 0.043 (0.036) | -0.199** (0.084) | -0.170** (0.079) | 0.025 (0.029) | -0.174** (0.070) | -0.151** (0.066) |
| (4) Constant | 0.999*** (0.010) | 0.359*** (0.050) | 0.360*** (0.050) | 0.996*** (0.006) | 0.436*** (0.034) | 0.431*** (0.033) | 0.997*** (0.009) | 0.426*** (0.035) | 0.423*** (0.035) | 0.991*** (0.008) | 0.377*** (0.033) | 0.372*** (0.033) | 0.999*** (0.007) | 0.354*** (0.028) | 0.352*** (0.027) |
| (5) Observations | 491 | 475 | 491 | 952 | 927 | 952 | 706 | 681 | 706 | 999 | 953 | 998 | 1414 | 1356 | 1414 |
| (6) Adjusted R-squared | 0.046 | 0.062 | 0.060 | 0.102 | 0.102 | 0.107 | 0.106 | 0.092 | 0.090 | 0.124 | 0.110 | 0.114 | 0.089 | 0.097 | 0.100 |

Notes: Table shows estimated effects on eyeglass ownership and usage relative to the Free Distribution without Information group. Standard errors in parentheses account for clustering at the school level. Conditional usage is estimated using the subsample owning eyeglasses. Each regression estimated using the subsample indicated above. *** Significant at the 1 percent level. Corresponding estimates using an index constructed using these variables is shown in Table 7.
*** Significant at the 1 percent level.
** Significant at the 5 percent level.
* Significant at the 10 percent level.